\documentclass[aps,pre,onecolumn,floatfix]{revtex4}

\usepackage{graphicx}
\usepackage{amssymb,amsfonts,amsmath}
\usepackage{epsf}
\usepackage{subfigure}
\usepackage{epstopdf}
\usepackage{mathrsfs}
\usepackage{color}
\usepackage{ulem}

\newcommand{\be}{\begin{equation}}
\newcommand{\ee}{\end{equation}}
\newcommand{\bea}{\begin{eqnarray}}
\newcommand{\eea}{\end{eqnarray}}

\begin{document}

\title{The stochastic motion of self-thermophoretic Janus particles}

\author{Pierre Gaspard}
\email{gaspard@ulb.ac.be}
\affiliation{Center for Nonlinear Phenomena and Complex Systems, Universit{\'e} Libre de Bruxelles (U.L.B.), Code Postal 231, Campus Plaine, B-1050 Brussels, Belgium}
\author{Raymond Kapral}
\email{r.kapral@utoronto.ca }
\affiliation{Chemical Physics Theory Group, Department of Chemistry, University of Toronto, Toronto, Ontario M5S 3H6, Canada}

\begin{abstract}
Langevin equations for the self-thermophoretic dynamics of Janus motors partially coated with an absorbing layer that is heated by a radiation field are presented. The derivation of these equations is based on fluctuating hydrodynamics and radiative heat transfer theory involving stochastic equations for bulk phases and surface processes that are consistent with microscopic reversibility. Expressions for the self-thermophoretic force and torque for arbitrary slip boundary conditions are obtained. The overdamped Langevin equations for the colloid displacement and radiative heat transfer provide expressions for the self-thermophoretic velocity and its reciprocal contribution where an external force can influence the radiative heat transfer. A nonequilibrium fluctuation formula is also derived and shows how the probability density of the Janus particle displacement and radiation energy transfer during the time interval $[0,t]$ are related to the mechanical and thermal affinities that characterize the nonequilibrium system state.
\end{abstract}

\maketitle

\section{Introduction}

The response of fluid mixtures to both concentration and temperature gradients is well known and is embodied in the linear laws for the particle and heat fluxes. In addition to direct effects involving the diffusion and heat conductivity coefficients, these fluxes contain cross Onsager coefficients that describe the thermal diffusion or Soret effect, where a   concentration flux is generated by a temperature gradient, and the reciprocal Dufour effect where a heat flux is produced by a concentration gradient~\cite{GM84}.  Analogous phenomena occur in colloidal suspensions subjected to temperature gradients. In such systems the movements of colloidal particles are attributed to thermophoresis where the temperature gradient on the surface of the colloidal particle induces a fluid slip velocity that is responsible for particle motion~\cite{DD74,A86,A89}. Because thermophoresis involves a delicate interplay among surface forces, surface transport processes and temperature gradients, aspects of the mechanism remain the subject of current research being carried out from a number of different perspectives~\cite{PP04,D04a,D04b,PP08,DB06,W10,LYR12,YR13,FK16,BFPE18,BBPE18}.

In colloidal motion driven by self-thermophoresis, typically a colloidal particle is partially coated by a material that absorbs radiation and produces an inhomogeneous temperature field in the vicinity of the particle. As for the case of an externally applied temperature gradient, the self-produced temperature gradient leads to a velocity slip on the colloid surface that is responsible for its self-propulsion. The self-thermophoretic propulsion of spherical Janus and other colloids has been studied experimentally~\cite{JYS10,BC15,KCC16} and by simulation~\cite{YR11,YWR14}. The full elucidation of the self-propulsion mechanism and its description also involve challenging issues that stem from the processes that occur on the surface of the colloid. Since the colloidal particles are small, often with micrometer or sub-micrometer linear dimensions, it is also important to take thermal fluctuations into account in complete treatments of the dynamics.

In this paper, we adopt an approach based on fluctuating hydrodynamics and radiative heat transfer theory in order to construct Langevin equations for the force and torque on a Janus colloid, in the presence of a radiation field that heats the partially coated surface of the particle and an external force and torque. Section~\ref{sec:fluct-hydro} presents the fluctuating Navier-Stokes and heat equations in the bulk phase, along with the stochastic equations at the surface that are consistent with microscopic reversibility. The Langevin equations for the force, torque, and heat transfer are given in Sec.~\ref{sec:FTandHT} and involve the thermophoretic force and torque that are responsible for the active motion of the Janus colloid. The corresponding coupled overdamped equations in Sec.~\ref{sec:coupledL} for the colloid position and radiative heat transfer contain the self-thermophoretic velocity and its reciprocal effect. This section also presents a nonequilibrium fluctuation formula for the joint probability of the colloid displacement and the radiation energy transfer. Numerical results on self-thermophoretic enhancement of diffusion and propulsion when the motor orientation is controlled by an external magnetic field, and the stalling of motor motion by application of an external force and external torque, are given in Sec.~\ref{sec:num}.  The summary and perspectives on this work are presented in Sec.~\ref{sec:conc}. Results for the thermophoretic velocity for a Janus colloid subject to an external temperature gradient are discussed in Appendix~\ref{app:therm-vel}, while a method for simulating the Langevin equation is given in Appendix~\ref{app:sim}.

\section{Fluctuating hydrodynamics and radiative heat transfer at an interface}
\label{sec:fluct-hydro}

We consider a Janus particle in a fluid and heated by thermal radiation.  The particle is solid, has a spherical shape, and is capped by a thin hemispherical layer of material that absorbs electromagnetic radiation.  The Janus particle moves in a fluid that also serves as a heat bath.  This system is subjected to thermal radiation at temperature $T_0$.  Both the fluid and the bulk of the Janus particle are assumed to be transparent to the electromagnetic radiation.  Consequently, radiation is only absorbed by the hemispherical cap of the Janus particle, which is locally heated, causing a temperature profile around the Janus particle.  The resulting component of the temperature gradient that is parallel to the fluid-solid interface induces a velocity slip between the solid particle and the fluid, which leads to propulsion of the Janus particle by thermophoresis.

In this mechanism, radiative heat transfer generates the temperature in the system, which is coupled to the fluid flow around the particle.  Therefore, the description of the processes involved in this mechanism requires the determination of the fluid velocity field from the Navier-Stokes equations and the temperature field from the heat equation, both being coupled by boundary conditions that account for thermophoresis.  The aim of this work is to obtain the Langevin equations that describe the stochastic motion of the Janus particle starting from fluctuating hydrodynamics that incorporates the effects of thermal fluctuations in the system, coupled to the radiative heat transfer from the external source of the thermal radiation.

\subsection{Stochastic equations for the bulk phases}

\subsubsection{The velocity field}

The fluid is assumed to be incompressible so that the velocity field $\bf v$ obeys $\pmb{\nabla}\cdot{\bf v}=0$ and the mass density $\rho$ is everywhere constant in time. In the fluid, the velocity field satisfies the fluctuating Navier-Stokes equations
\be\label{NS-eqs}
\rho\left(\partial_t{\bf v} + {\bf v}\cdot\pmb{\nabla}{\bf v}\right) = -\pmb{\nabla}\cdot{\boldsymbol{\mathsf P}} \, ,
\ee
where the pressure tensor is expressed as ${\boldsymbol{\mathsf P}}=P{\boldsymbol{\mathsf 1}}+\pmb{\Pi}$ in terms of the hydrostatic pressure $P$ and the viscous pressure tensor \cite{LL80Part2,G04,OS06,BM74}
\be
\Pi_{ij} = -\eta \left( \partial_i v_j+\partial_j v_i\right) + \pi_{ij} \, ,
\label{Pi}
\ee
where $\eta$ is the shear viscosity and $\pi_{ij}$ are Gaussian white noise fields characterized by
\be \label{pi-correl}
\langle \pi_{ij}({\bf r},t)\rangle = 0 \qquad\mbox{and}\qquad\langle \pi_{ij}({\bf r},t)\, \pi_{kl}({\bf r}',t')\rangle = 2 k_{\rm B}T \eta \left(\delta_{ik}\delta_{jl}+\delta_{il}\delta_{jk}\right)\delta({\bf r}-{\bf r}') \, \delta(t-t')\, ,
\ee
where $k_{\rm B}$ is Boltzmann's constant.

The radius of the Janus particle is denoted by $R$.  The position of the Janus particle center of mass, its translational velocity, and its angular velocity are respectively denoted ${\bf R}(t)$, ${\bf V}(t)$, and $\pmb{\Omega}(t)$. Thus, inside the solid Janus particle for $\Vert{\bf r}-{\bf R}(t)\Vert <R$, the velocity field is given by \cite{MB74}
\be\label{v-sol}
{\bf v}({\bf r},t) = {\bf V}(t) + \pmb{\Omega}(t)\times \left[ {\bf r}-{\bf R}(t)\right] .
\ee

\subsubsection{The temperature field}

The heat equation for the temperature field $T$ has the form
\be\label{heat-eq}
\rho c_V \left(\partial_t T + {\bf v}\cdot\pmb{\nabla}T\right) = -\pmb{\nabla}\cdot{\bf J}_q - \pmb{\Pi}:\pmb{\nabla}{\bf v} + \phi\, ,
\ee
expressed in terms of the specific heat capacity $c_V$ and the heat current density given by Fourier's law
\be
{\bf J}_q = -\kappa\, \pmb{\nabla} T + \pmb{\eta}_q \, ,
\label{Fourier}
\ee
where $\kappa$ is the heat conductivity and $\pmb{\eta}_q$ is a vector of Gaussian white noise fields obeying
\be \label{eta-correl}
\langle \eta_{qi}({\bf r},t)\rangle = 0 \qquad\mbox{and}\qquad\langle  \eta_{qi}({\bf r},t)\,  \eta_{qj}({\bf r}',t')\rangle = 2 k_{\rm B}T^2 \kappa\, \delta_{ij} \delta({\bf r}-{\bf r}') \, \delta(t-t')\, ,
\ee
and $\phi={\bf j}_{\rm e}\cdot\pmb{\cal E}$  is the heating power density of radiation due to the effect of the time-dependent electric field $\pmb{\cal E}$ on the electric current density ${\bf j}_{\rm e}$.
We note that the term $-\pmb{\Pi}:\pmb{\nabla}{\bf v}$ in the right hand side of Eq.~(\ref{heat-eq}) describes heating due to shear viscosity in the fluid, but it is vanishing in the solid particle.  The heat conductivity is denoted $\kappa^+$ for  the fluid and $\kappa^-$ for  the solid.

\subsection{Stochastic equations at the interface}

\subsubsection{Interfacial radiative heat transfer}

Since the bulk solid and fluid phases are transparent to radiation, heating by radiation is concentrated on the thin absorbing layer of the Janus particle, so that the heating power density is given by $\phi=\phi^{\rm s}\,\delta^{\rm s}({\bf r},t)$, where $\phi^{\rm s}$ has the units of a power per unit area and $\delta^{\rm s}({\bf r},t)$ is the interfacial Dirac distribution \cite{BAM76}.  According to the Stefan-Boltzmann law of grey-body thermal radiation, the net radiative heating power per unit area is given by
\be
\phi^{\rm s} = \epsilon \sigma (T_0^4-T^4) + \xi_{\phi}^{\rm s} \, ,
\label{phi-s}
\ee
where $\epsilon<1$ is the dimensionless emissivity of the absorbing layer, $\sigma=5.67\times 10^{-8}$~W/(m$^2$ K$^4$) the Stefan-Boltzmann constant, $T_0$ the temperature of the black-body radiation source, and $T$ the local temperature of the thin absorbing layer. The quantity $\phi^{\rm s}$ has the units W/m$^2$. Since heating by radiation is a stochastic process, the added Gaussian white noise term $\xi_{\phi}^{\rm s}$ satisfies
\be
\langle\xi_{\phi}^{\rm s}({\bf r},t)\rangle = 0 \qquad\mbox{and} \qquad \delta^{\rm s}({\bf r},t)\, \langle\xi_{\phi}^{\rm s}({\bf r},t)\, \xi_{\phi}^{\rm s}({\bf r}',t')\rangle\, \delta^{\rm s}({\bf r}',t')= 2k_{\rm B} L_{\phi\phi}^{\rm s}\,   \delta^{\rm s}({\bf r},t) \, \delta({\bf r}-{\bf r}') \, \delta(t-t') \, ,
\label{phi-noise}
\ee
where
\be
L_{\phi\phi}^{\rm s} = 2 \epsilon\sigma (T_0^5+T^5)
\label{L-phi-phi}
\ee
is the coefficient determining the interfacial diffusivity of radiation that is absorbed and emitted by the thin absorbing layer.  This coefficient is obtained as follows.  In the same way as the fluctuations of energy $\hat U$ have a variance given by ${\rm Var}(\hat U)=k_{\rm B} T^2 \partial\langle\hat U\rangle/\partial T$ in the Gibbsian canonical equilibrium ensemble, the fluctuations of the energy radiated by a black-body per unit time and unit area has mean value $\langle\hat R\rangle=\sigma T^4$ and variance ${\rm Var}(\hat R)=k_{\rm B} T^2 \partial\langle\hat R\rangle/\partial T = 4k_{\rm B}\sigma T^5$.  For a grey-body, this quantity should be multiplied by the emissivity.  Moreover, the variances of the absorbed and emitted thermal radiation energies should be added.  Now, the net energy $\hat E$ transferred from radiation to an area $\Delta\Sigma$ of the thin absorbing layer during the time interval $\Delta t$ is given by $\Delta \hat E=\int_{\Delta t} dt \int_{\Delta\Sigma}d\Sigma\, \phi^{\rm s}$.  According to Eqs.~(\ref{phi-s}) and~(\ref{phi-noise}), the mean value and the variance of this net fluctuating energy are given by $\langle\Delta\hat E\rangle=\epsilon \sigma (T_0^4-T^4)\Delta t \Delta\Sigma$ and ${\rm Var}(\Delta\hat E)=2k_{\rm B}L_{\phi\phi}^{\rm s}\Delta t \Delta\Sigma=2 \epsilon\sigma (T_0^5+T^5)\Delta t \Delta\Sigma$, in accord with Eq.~(\ref{L-phi-phi}).

\subsubsection{Temperature-velocity coupling at the fluid-solid interface}

In the following, any field $a$ is denoted $a^+$ in the fluid and $a^-$ in the solid.

The velocity field ${\bf v}^{+}$ in the fluid obeys the fluctuating Navier-Stokes equations~(\ref{NS-eqs}) and the velocity field ${\bf v}^{-}$ is given by Eq.~(\ref{v-sol}).  If $\bf n$ denotes a unit vector normal to the interface, the normal component of the velocity field obeys the boundary condition
\be
{\bf n}\cdot{\bf v}^{+}({\bf r},t) = {\bf n}\cdot{\bf v}^{-}({\bf r},t) = {\bf n}\cdot{\bf V}(t)
\ee
for $\Vert{\bf r}-{\bf R}(t)\Vert=R$.  This boundary condition does not determine the velocity slip,
\be
{\bf v}_{\rm slip} \equiv {\bf v}^{+}-{\bf v}^{-} \, ,
\ee
which is oriented in the direction parallel to the interface and is coupled to the tangential gradient of temperature by thermophoresis.  The thermophoretic coupling is established by interfacial nonequilibrium thermodynamics \cite{W67,BAM76,K77,B86} according to the following phenomenological linear relations:
\bea
{\bf n}\cdot{\pmb{\Pi}}\cdot{\boldsymbol{\mathsf 1}}_{\bot} &=& -\frac{L^{\rm s}_{\rm vv}}{T}\, {\bf v}_{\rm slip} - \frac{L^{\rm s}_{{\rm v}q}}{T^2}\, \pmb{\nabla}_{\bot}T + \pmb{f}^{\rm s}_{\rm fl}\,  ,  \qquad \label{Pi-s+n}\\
{\bf J}_q^{\rm s} &=& -\frac{L^{\rm s}_{q{\rm v}}}{T}\, {\bf v}_{\rm slip}  - \frac{L^{\rm s}_{qq}}{T^2}\, \pmb{\nabla}_{\bot}T+ \pmb{\eta}_q^{\rm s}\, , \label{Jq-s}
\eea
where ${\boldsymbol{\mathsf 1}}_{\bot}\equiv{\boldsymbol{\mathsf 1}}-{\bf n}{\bf n}$, $\pmb{\nabla}_{\bot}$ is the tangential gradient, $T=T^{+}$, and $L_{ij}^{\rm s}$ are linear response coefficients coupling the affinities or thermodynamic forces $-{\bf v}_{\rm slip}/T$ and $-\pmb{\nabla}_{\bot}T/T^2$ to their respective interfacial currents ${\bf n}\cdot{\pmb{\Pi}}\cdot{\boldsymbol{\mathsf 1}}_{\bot}$ for interfacial momentum and ${\bf J}_q^{\rm s}$ for interfacial heat.  In the notations of Ref.~\cite{BAM76}, the linear response coefficients read $L^{\rm s}_{\rm vv} = T  \, \beta$, $L^{\rm s}_{qq}= T^2 \, \lambda^{\rm s}$, $L^{\rm s}_{q{\rm v}} = T\alpha_{12}$, and  $L^{\rm s}_{{\rm v}q} = T\alpha_{21}$.

Since the affinity $-{\bf v}_{\rm slip}/T$ is odd under time reversal while the other affinity $-\pmb{\nabla}_{\bot}T/T^2$ is even, microreversibility implies the Onsager-Casimir reciprocal relation
\be
L^{\rm s}_{{\rm v}q} = - L^{\rm s}_{q{\rm v}}
\label{L-L}
\ee
between the coefficient $L^{\rm s}_{{\rm v}q}$ of thermophoresis and the coefficient $L^{\rm s}_{q{\rm v}}$ describing the reciprocal effect of the velocity slip back onto the interfacial heat current density ${\bf J}_q^{\rm s}$.  Accordingly, the interfacial Gaussian white noise processes satisfy $\langle\pmb{f}_{\rm fl}^{\rm s}({\bf r},t)\rangle = 0$, $\langle{\pmb{\eta}}_q^{\rm s}({\bf r},t)\rangle = 0$, and
\bea
&& \delta^{\rm s}({\bf r},t)\, \langle\pmb{f}_{\rm fl}^{\rm s}({\bf r},t)\, \pmb{f}_{\rm fl}^{\rm s}({\bf r}',t')\rangle\, \delta^{\rm s}({\bf r}',t') = 2 k_{\rm B}L^{\rm s}_{\rm vv}\, \delta^{\rm s}({\bf r},t) \, \delta({\bf r}-{\bf r}') \, \delta(t-t') \, {\boldsymbol{\mathsf 1}}_{\bot} \, ,
\label{f-f} \\
&& \delta^{\rm s}({\bf r},t)\, \langle\pmb{f}_{\rm fl}^{\rm s}({\bf r},t)\, {\pmb{\eta}}_q^{\rm s}({\bf r}',t')\rangle\, \delta^{\rm s}({\bf r}',t') = 0 \, ,
\label{f-eta0} \\
&& \delta^{\rm s}({\bf r},t)\, \langle{\pmb{\eta}}_q^{\rm s}({\bf r},t)\, {\pmb{\eta}}_q^{\rm s}({\bf r}',t')\rangle\, \delta^{\rm s}({\bf r}',t') = 2 k_{\rm B}L^{\rm s}_{qq}\, \delta^{\rm s}({\bf r},t) \, \delta({\bf r}-{\bf r}') \, \delta(t-t') \, {\boldsymbol{\mathsf 1}}_{\bot} \, ,
\label{eta0-eta0}
\eea
without coupling between the noise fields because of the antisymmetry~(\ref{L-L}) of the Onsager-Casimir reciprocal relation.  The linear response coefficients can be written as
\be
L^{\rm s}_{\rm vv} = T  \, \lambda, \quad L^{\rm s}_{qq}= T^2 \, \kappa^{\rm s}, \quad L^{\rm s}_{{\rm v}q} = T^2 \, \lambda\,  b_q \, ,
\ee
in terms of the coefficient of sliding friction $\lambda$ of Ref.~\cite{BB13}, the interfacial heat conductivity $\kappa^{\rm s}$, and the thermophoretic constant $b_q$.  Moreover, the slip length is defined as $b=\eta/\lambda$. Accordingly, the boundary condition for the tangential component of the velocity field is given by
\bea
{\bf v}_{\rm slip} &\equiv& {\boldsymbol{\mathsf 1}}_{\bot}\cdot\left\{ {\bf v}({\bf r},t) -{\bf V}(t) - \pmb{\Omega}(t)\times\left[{\bf r}-{\bf R}(t)\right]\right\} \nonumber\\
&=&{\boldsymbol{\mathsf 1}}_{\bot}\cdot\biggl\{ - b_q \pmb{\nabla}T({\bf r},t) + b\left[\pmb{\nabla}{\bf v}({\bf r},t) +\pmb{\nabla}{\bf v}({\bf r},t)^{\rm T}\right]\cdot{\bf n} +\pmb{f}^{\rm s}_{\rm fl}({\bf r},t)/\lambda\biggr\} \label{v-bc2}
\eea
for $\Vert{\bf r}-{\bf R}(t)\Vert =R$, where ${\bf v}={\bf v}^+$ and $T=T^{+}$.

\subsubsection{Temperature boundary condition at the fluid-solid interface}

In the framework of interfacial nonequilibrium thermodynamics \cite{BAM76}, the boundary condition on the temperature field is determined by considering the balance equation for the excess surface density $e^{\rm s}$ of energy at the interface:
\be\label{eq-e-s}
\partial_t e^{\rm s} + \pmb{\nabla}_{\bot}\cdot\left( e^{\rm s}\, {\bf v}^{\rm s} + {\boldsymbol{\mathsf P}}^{\rm s}\cdot{\bf v}^{\rm s} +{\bf J}_q^{\rm s} \right) = -{\bf n}\cdot\left( {\boldsymbol{\mathsf P}}^+\cdot{\bf v}^{+} - {\boldsymbol{\mathsf P}}^-\cdot{\bf v}^{-} \right)-{\bf n}\cdot\left( {\bf J}_q^+ - {\bf J}_q^-\right) +\phi^{\rm s} \, ,
\ee
where ${\bf v}^{\rm s}=({\bf v}^++{\bf v}^-)/2$ is the interfacial velocity and ${\bf J}_q^{\rm s}$ is the heat surface current density given by Eq.~(\ref{Jq-s}).  Since the heat current densities in the fluid and the solid, ${\bf J}_q^{\pm}$, are given by Eq.~(\ref{Fourier}), Eq.~(\ref{eq-e-s}) constitutes a boundary condition on the temperature fields $T^{\pm}$ across the interface.  If changes in the excess surface energy density and heating due to sliding friction are assumed to be negligible, the standard boundary condition is recovered for the temperature field: ${\bf n}\cdot\left( {\bf J}_q^+ - {\bf J}_q^-\right) = \phi^{\rm s}$ where ${\bf J}_q^{\pm}$ are the heat current densities~(\ref{Fourier}) on both sides of the interface.

\subsubsection{The thermophoretic constant in the thin-layer approximation}

According to Refs.~\cite{DCM87,GLF17,FMJ17}, the thermophoretic constant $b_q$ is given by
\be
b_q = \frac{1}{\eta T} \left( H^{(1)}+b H^{(0)}\right) \qquad\mbox{with}\quad H^{(n)}=\int_0^{\delta} dz \, z^n \, \Delta h(z) \, ,
\label{b_q}
\ee
for $n=0,1$, where $\Delta h(z)=h(z)-h_{\rm bulk}$ is the excess enthalpy density at height $z$ above the surface, defined as the difference between the local enthalpy $h(z)$ modified due to the interaction of the fluid with the surface through some interaction potential $u(z)$, and the bulk value of the enthalpy density $h_{\rm bulk}$.  In the special case where the fluid is a dilute gas, the molecular density behaves as $n(z)\sim\exp[-\beta u(z)]$, so that we have
\be
H^{(n)} = h_{\rm bulk} \int_0^{\delta} dz \, z^n \, \left[{\rm e}^{-\beta u(z)}-1\right] ,
\ee
showing the analogy with the diffusiophoretic constant \cite{A89}.

\section{Force, torque, and heat transfer rate}\label{sec:FTandHT}

\subsection{The motion of the Janus particle}

In this section we present equations of motion for the Janus particle due to its interaction with the surrounding fluid and heating by radiation of both the particle and the fluid.  Through thermophoresis, heating determines the slip velocity between the particle and the fluid. In turn, the fluid exerts a force at the surface $\Sigma(t)$ of the particle. The techniques used to obtain the stochastic equations of motion closely parallel those for diffusiophoresis so the presentation given below will focus on new aspects that arise in the treatment of thermophoresis.

The force on the Janus particle can be expressed in terms of the surface integral of the fluid pressure tensor entering Navier-Stokes equations~(\ref{NS-eqs}). Including the presence of an external force  ${\bf F}_{\rm ext}$, Newton's equation for the Janus particle reads
\be
M\, \frac{d{\bf V}}{dt} = -\int_{\Sigma(t)} {\boldsymbol{\mathsf P}}({\bf r},t)\cdot{\bf n} \, d\Sigma + {\bf F}_{\rm ext},
\label{Langevin-Eq-0}
\ee
where $M=\int_{{\cal V}(t)}\rho_{\rm solid} \, d{\bf r}$ is the mass of the Janus particle, $\rho_{\rm solid}$ being its mass density \citep{MB74,BM74}. In addition, the fluid exerts a torque on the Janus particle and its angular velocity obeys the equation,
\be
{\boldsymbol{\mathsf I}}\cdot\frac{d\pmb{\Omega}}{dt} = -\int_{\Sigma(t)} \Delta{\bf r}\times \left[{\boldsymbol{\mathsf P}}({\bf r},t)\cdot{\bf n}\right]  d\Sigma + {\bf T}_{\rm ext},
\label{rot-Langevin-Eq-0}
\ee
where the inertia tensor ${\boldsymbol{\mathsf I}}$ of the Janus particle has the components $I_{ij}=\int_{{\cal V}(t)}\rho_{\rm solid} \big(\Delta{\bf r}^2\, \delta_{ij}-\Delta r_i\Delta r_j\big) d{\bf r}$ with $\Delta{\bf r}\equiv{\bf r}-{\bf R}(t)$, and ${\bf T}_{\rm ext}$ is an external torque \citep{H75,F76a,F76b,BAM77}. Furthermore, the Janus particle is driven out of equilibrium, not only by the external force ${\bf F}_{\rm ext}$, but also by the radiative heat transfer.  Accordingly, the above equations of motion are coupled to the overall radiative heat transfer given by
\be
\Phi= \int_{\Sigma(t)} \phi^{\rm s} \, d\Sigma \, ,
\label{Phi}
\ee
where the surface integral is carried out over the absorbing hemisphere of the Janus particle.

In order to obtain the right sides of these equations, the fluid velocity field and the temperature field should be determined by solving the partial differential equations in Sec.~\ref{sec:fluct-hydro}.  For this purpose, it is useful to consider the different time scales involved in the motion of the Janus particle, as given in Table~\ref{table1}.  These time scales should be compared with the observational time scale, which is of the order of seconds.

\begin{table}[h]
\caption{Time scales of a spherical Janus particle of radius $R=1\, \mu$m mostly composed of silica ($\rho_{\rm solid}=2200$~kg/m$^3$) moving in water at the temperature $20^{\circ}$C, with the self-thermophoretic velocity $V_{\rm sth}=10\, \mu$m/s and stick boundary conditions $(b=0$), together with the time scales of the velocity and temperature fields around it. The viscosity of water is about $\eta\simeq 10^{-3}\, {\rm N\, s}/{\rm m}^2$, its kinematic viscosity $\nu\simeq 10^{-6}\, {\rm m}^2/{\rm s}$, its thermal conductivity $\kappa\simeq 0.6\times 10^{-2} \, {\rm W}/{\rm m \, K}$, its thermal diffusivity $D_{\rm th}\simeq 1.4\times 10^{-7} \, {\rm m}^2/{\rm s}$, and its Prandtl number ${\rm Pr}=\nu/D_{\rm th}\simeq 7$.}
{\begin{tabular}{ll}
\hline\hline
 time scale & order of magnitude \\
\hline
sound & $t_{\rm sound}\sim \frac{R}{v_{\rm sound}}\simeq 7\times 10^{-10}$~s  \\
angular velocity thermalization & $t_{\rm rot\; therm}=\frac{I}{\gamma_{\rm r}}=\frac{R^2}{15\eta}\rho_{\rm solid}\simeq 1.5\times 10^{-7}$~s \\
velocity thermalization & $t_{\rm transl\; therm}=\frac{M}{\gamma_{\rm t}}=\frac{2R^2}{9\eta}\rho_{\rm solid}\simeq 5\times 10^{-7}$~s \\
hydrodynamics & $t_{\rm hydro}\sim \frac{R^2}{\eta}\rho_{\rm fluid}\simeq 10^{-6}$~s \\
thermal diffusivity & $t_{\rm th \; diff}\sim \frac{R^2}{D_{\rm th}}\simeq 7\times 10^{-6}$~s  \\
propulsion & $t_{\rm prop}=\frac{R}{V_{\rm sth}}\simeq 10^{-1}$~s   \\
rotational diffusion & $t_{\rm rot \; diff}= \frac{1}{2D_{\rm r}}\simeq 3$~s  \\
translational diffusion & $t_{\rm transl\; diff}\sim \frac{R^2}{D_{\rm t}}\simeq 5$~s  \\
\hline
\end{tabular}}
\label{table1}
\end{table}

The shortest time scale is associated with sound, below which the effects of fluid compressibility become observable.  Therefore, the system may be considered as incompressible.  The time scales of thermalization for the translational and rotational velocities of the Janus particle mark the crossover for the importance of the inertial effects in the movement of the Janus particle.  For $t\gg t_{\rm rot\; therm}$ and $t\gg t_{\rm transl\; therm}$, the Langevin equations describing the random motion of the Janus particle may be considered as overdamped.  For $t\gg t_{\rm hydro}$, the memory effects of the fluid flow around the Janus particle become negligible.  In this regime, the translational and rotational friction coefficients become time independent and they are respectively given by
\be \label{t-friction}
\gamma_{\rm t}= 6\pi\eta R \, \frac{1+2b/R}{1+3b/R},
\ee
and
\be \label{r-friction}
\gamma_{\rm r}=\frac{8\pi\eta  R^3}{1+3b/R},
\ee
for a spherical particle of radius $R$ and hydrophobicity characterized by the slip length $b$.

Under constant radiation heating, the temperature profile becomes stationary around the Janus particle on time scales $t\gg t_{\rm th\; diff}$, where $t_{\rm th\; diff}$ depends on the thermal diffusivity $D_{\rm th}=\kappa/(\rho c_V)$.  In water at $20^\circ$C, the thermal diffusivity takes the value $D_{\rm th}\simeq 1.4\times 10^{-7}$~m$^2$/s, giving $t_{\rm th\; diff}\sim R^2/D_{\rm th}\simeq 7\times 10^{-6}$~s.  In silica, we have $D_{\rm th}\simeq 9\times 10^{-7}$~m$^2$/s, and thus $t_{\rm th\; diff}\sim R^2/D_{\rm th}\simeq 10^{-6}$~s. We note that this time scale is significantly shorter than that associated with molecular diffusivity $t_{\rm mol \; diff}\sim R^2/D_{\rm mol}\simeq 10^{-3}$~s, since typically $D_{\rm mol}\simeq 10^{-9}$~m$^2$/s.  It is over this time scale that the concentration profiles become stationary around Janus particles in self-diffusiophoresis.  Therefore, the assumption that the temperature profile remains stationary in the frame accompanying the Janus particle in all its translational and rotational movements is very well satisfied, because the time scales of propulsion, rotational diffusion, and translational diffusion are much longer.

The fluid flow is laminar for micrometric particles moving at a self-thermophoretic velocity $V_{\rm sth}\simeq 10\, \mu$m/s in an aqueous solution, because the Reynolds number is much smaller than unity, ${\rm Re}\equiv V_{\rm sth}R/\nu \sim 10^{-5}$.  In this regime, the Navier-Stokes equations~(\ref{NS-eqs}) can be linearized by neglecting all the terms that are nonlinear in the velocity field.  For the same reason, the dependence of viscosity on temperature is also neglected. Since the Prandtl number of water is ${\rm Pr}=\nu/D_{\rm th}\simeq 7$, the thermal P\'eclet number is also smaller than unity, ${\rm Pe}_{\rm th}\equiv V_{\rm sth}R/D_{\rm th} = {\rm Re}\; {\rm Pr} \sim 10^{-4}$.  Consequently, the advective term an be neglected in the heat equation~(\ref{heat-eq}).  Moreover, the heating power due to fluid shear viscosity is also negligible in the laminar regime because it is quadratic in the fluid velocity gradients in Eq.~(\ref{heat-eq}).  Therefore, the heat equation can also be linearized.

\subsection{Langevin equations}

By solving the linear Navier-Stokes equations, we obtain the Langevin equations for the translational and rotational movements of the spherical Janus particle:
\bea
M\frac{d{\bf V}}{dt} = -\gamma_{\rm t}\, {\bf V}  + {\bf F}_{\rm th} + {\bf F}_{\rm ext} + {\bf F}_{\rm fl}(t) \, ,
\label{Langevin-eq-transl} \\
{\boldsymbol{\mathsf I}}\cdot\frac{d\pmb{\Omega}}{dt} = - \gamma_{\rm r}\,\pmb{\Omega} + {\bf T}_{\rm th} + {\bf T}_{\rm ext} + {\bf T}_{\rm fl}(t) \, ,
\label{Langevin-eq-rot}
\eea
in terms of the friction coefficients~(\ref{t-friction}) and~(\ref{r-friction}), and the external and fluctuating forces, ${\bf F}_{\rm ext}$ and ${\bf F}_{\rm fl}$, and torques, ${\bf T}_{\rm ext}$ and ${\bf T}_{\rm fl}$, respectively.  The force ${\bf F}_{\rm th}$ and torque ${\bf T}_{\rm th}$ due to thermophoresis find their origin in the boundary condition~(\ref{v-bc2}) on the tangential component of the velocity field.  Around a spherical particle, this boundary condition can be taken into account with methods developed in Refs.~\cite{MB74,BM74,ABM75} and already applied to diffusiophoresis in Refs.~\cite{GK17,GK18a,GK18c}.  In this way, we find the following expressions:
\be
{\bf F}_{\rm th} = \frac{6\pi\eta R}{1+3b/R} \, \overline{b_q\, {\boldsymbol{\mathsf 1}}_{\bot}\cdot\pmb{\nabla}T({\bf r},t)}^{\rm s} ,
\label{Fth}
\ee
and
\be
{\bf T}_{{\rm th}}= \frac{12\pi\eta  R}{1+3b/R} \, \overline{b_q\, {\bf r}\times\pmb{\nabla}T({\bf r},t)}^{\rm s} ,
\label{Tth}
\ee
where $\overline{(\cdot)}^{\rm s} =(4\pi R^2)^{-1} \int (\cdot ) \, d\Sigma$ denotes the average over the surface of the Janus particle.
The thermophoretic translational and angular velocities are thus given by
\be
{\bf V}_{\rm th} =\frac{{\bf F}_{\rm th}}{\gamma_{\rm t}} = \frac{1}{1+2b/R} \, \overline{b_q\, {\boldsymbol{\mathsf 1}}_{\bot}\cdot\pmb{\nabla}T({\bf r},t)}^{\rm s} ,
\label{V-th}
\ee
and
\be
\pmb{\Omega}_{{\rm th}}=\frac{{\bf T}_{{\rm th}}}{\gamma_{\rm r}}= \frac{3}{2R^2} \, \overline{b_q\, {\bf r}\times\pmb{\nabla}T({\bf r},t)}^{\rm s}.
\label{Omega-th}
\ee
Accordingly, the effects of thermophoresis are determined by the temperature profile at the surface of the Janus particle.

\subsection{Temperature profile around the heated Janus particle}

The temperature field around the Janus particle can be obtained using the results of Refs.~\cite{JYS10,BMW13,BZW14}.
Since the Janus particle is axisymmetric, the reference frame used to solve the heat equation can be chosen with its $z$-axis coinciding with the symmetry axis of the particle and its origin at the center of the particle.  The surface of the particle is at the radius $r=R$ and the absorbing layer on the hemisphere with $z>0$. The stationary temperature field is thus obtained by solving the following equations:
\bea
&&\nabla^2 T^{\pm}(r,\theta)= 0 \, , \label{Laplace-eq}\\
&&T^+(R,\theta) = T^-(R,\theta) \, , \label{T-bc1}\\
&&-\kappa^+\partial_rT^+(R,\theta) +\kappa^- \partial_rT^-(R,\theta) = \phi^{\rm s}(\theta) \, , \label{T-bc2}\\
&&\lim_{r\to\infty} T^+(r,\theta) = T_{\infty} \, , \label{T-bc3}
\eea
written in spherical coordinates $(r,\theta,\varphi)$.  Because of the axial symmetry, the problem is independent of the angle $\varphi$.  The deterministic part of the temperature profile is obtained by neglecting the fluctuations in Eq.~(\ref{phi-s}).  Moreover, the variations of local temperature are usually very small with respect to the temperatures $T_0$ of the radiation source and $T_{\infty}$ of the fluid at large distances from the particle.  Therefore, the radiative heating power per unit area in Eq.~(\ref{T-bc2}) is taken as
\be
\phi^{\rm s}(\theta) = \epsilon\, I \, {\cal H}(\cos\theta), \qquad {\rm with \; radiation \; intensity} \quad I = \sigma \left( T_0^4-T_{\infty}^4\right) ,
\label{intensity}
\ee
and the Heaviside function ${\cal H}(\xi)=1$ if $\xi>0$ and ${\cal H}(\xi)=0$ if $\xi<0$, describing the absorption of radiation on the capped hemisphere.

The heating power can be expanded in a series of Legendre polynomials as $\phi^{\rm s}(\theta) = \sum_{l=0}^{\infty} \phi^{\rm s}_l \, P_l(\cos\theta)$ with coefficients $\phi^{\rm s}_l = \epsilon I  (l+1/2) \int_0^1 P_l(\xi) d\xi$.  The first few coefficients are given by $\phi^{\rm s}_0 = \epsilon I/2$, $\phi^{\rm s}_1 =  3\epsilon I/4$, $\phi^{\rm s}_3 =  -7 \epsilon I/16, \dots$ and $\phi^{\rm s}_l=0$ if $l$ is even.

The temperature field is thus given by
\bea
&& r>R: \qquad T^+(r,\theta) = T_{\infty} + \sum_{l=0}^{\infty} \tau_l \, P_l(\cos\theta)  \, \left(\frac{R}{r}\right)^{l+1}  , \\
&& r<R: \qquad T^-(r,\theta) = T_{\infty} + \sum_{l=0}^{\infty} \tau_l \, P_l(\cos\theta) \, \left(\frac{r}{R}\right)^l  ,
\eea
where the expansion coefficients take the values, $\tau_l = R \, \phi_l^{\rm s}/[\kappa^+(l+1)+\kappa^- l]$.
At the surface of the Janus particle, the temperature has thus the profile
\be
T^+(R,\theta) = T^-(R,\theta) = T_{\infty} + \sum_{l=0}^{\infty} \tau_l \, P_l(\cos\theta) \, .
\label{T-surf}
\ee
As a consequence, the mean value of the radiative heating rate is equal to
\be
\Phi_E = R^2 \int \phi^{\rm s}(\theta) \, d\cos\theta \, d\varphi = 4\pi R^2 \phi_0^{\rm s} = 2\pi R^2 \epsilon \, I \, .
\label{Phi-E}
\ee

\subsection{Self-thermophoretic velocity}

The self-thermophoretic velocity can be obtained from Eq.~(\ref{V-th}) using the interfacial temperature profile~(\ref{T-surf}).  Because of the axial symmetry, the velocity is always oriented in the direction of the unit vector $\bf u$ along this symmetry axis and pointing towards the absorbing hemisphere, i.e., the $z$-axis in the reference frame of previous subsection:
\be
{\bf V}_{\rm sth} = V_{\rm sth} \, {\bf u} \, .
\label{Vsth-u}
\ee

If we suppose that the thermophoretic constant takes the value $b_q^C$ on the capped hemisphere and the value $b_q^N$ on the other hemisphere, the magnitude of the self-thermophoretic velocity is found to be
\be
V_{\rm sth} = \frac{1}{1+2b/R} \, \sum_{l=0}^{\infty} \frac{\phi_l^{\rm s}}{\kappa^+(l+1)+\kappa^- l} \left( b_q^C \, \Delta_l^C + b_q^N \, \Delta_l^N \right) ,
\label{Vsth-2}
\ee
where the $\Delta_l^h$ quantities are defined by the integrals
\be
\Delta_l^h = -\frac{1}{2} \int_0^{\pi} d\theta \, \sin^2\theta \, {\cal H}_h(\cos\theta) \, \partial_{\theta} P_l(\cos\theta)
\ee
in terms of the Heaviside function ${\cal H}_h(\cos\theta)$, indicating the $h=C,N$ hemisphere.

In the case where the thermophoretic constant is uniform on the entire sphere $b_q^C=b_q^N=b_q$, the self-thermophoretic velocity is given by
\be
V_{\rm sth} = \frac{2\, b_q}{3(1+2b/R)} \, \frac{\phi_1^{\rm s}}{2\kappa^++\kappa^-} = \frac{b_q}{2(1+2b/R)} \, \frac{\epsilon \, I }{2\kappa^++\kappa^-} .
\label{Vsth}
\ee
We see that both the mean heating power~(\ref{Phi-E}) and the self-thermophoretic velocity~(\ref{Vsth}) are proportional to the radiation intensity $I$ in Eq.~(\ref{intensity}).  According they both vanish at thermodynamic equilibrium.  A self-thermophoretic proportionality parameter can be introduced
\be
\chi = \frac{V_{\rm sth} }{\Phi_E} = \frac{b_q}{4\pi R^2 (1+2b/R)(2\kappa^++\kappa^-)} ,
\label{chi}
\ee
which depends only on the material properties that are intrinsic to the Janus particle.

In addition, for an axisymmetric Janus particle, the self-thermophoretic angular velocity is equal to zero, $\pmb{\Omega}_{\rm sth}=0$, because the thermophoretic torque~(\ref{Tth}) vanishes in this case.

The expression for the self-thermophoretic velocity~(\ref{Vsth}) can be compared to the usual thermophoretic velocity of a spherical particle in a temperature gradient, which is given in Appendix~\ref{app:therm-vel}.

\section{Coupled Langevin equations in the overdamped regime}\label{sec:coupledL}

\subsection{Translation and rotation}

In the overdamped limit, the inertial term proportional to the mass $M$ is negligible in the Langevin equation~(\ref{Langevin-eq-transl}), which reduces to
\be
\frac{d{\bf R}}{dt} = {\bf V}_{\rm sth}  + \beta D_{\rm t} \, {\bf F}_{\rm ext} + {\bf V}_{\rm fl}(t) \, ,
\label{eq-R}
\ee
in terms of the self-thermophoretic velocity~(\ref{Vsth-u}), the inverse temperature $\beta=(k_{\rm B}T_{\infty})^{-1}$, the translational diffusion coefficient of the particle given by Einstein's formula $D_{\rm t}=(\beta\gamma_{\rm t})^{-1}$, and the fluctuating velocity ${\bf V}_{\rm fl}(t) ={\bf F}_{\rm fl}(t)/\gamma_{\rm t}$.  The self-thermophoretic velocity can be written as ${\bf V}_{\rm sth}=V_{\rm sth} {\bf u}=\chi \Phi_E {\bf u}$ in terms of the unit vector $\bf u$, the mean heating rate~(\ref{Phi-E}), and the thermophoretic parameter~(\ref{chi}).  The thermophoretic force and the parameter $\chi$ remain finite in the limits of perfect stick ($b\to 0$) and perfect slip ($b\to\infty$) boundary conditions.

The inertial term in the Langevin equation~(\ref{Langevin-eq-rot}) is also negligible in the overdamped limit.  Since the self-thermophoretic torque is zero, the directional unit vector $\bf u$ obeys
\be
\frac{d{\bf u}}{dt} = \left[\pmb{\Omega}_{\rm ext} + \pmb{\Omega}_{\rm fl}(t) \right] \times {\bf u} \, ,
\label{eq-rot}
\ee
where $\pmb{\Omega}_{\rm ext}={\bf T}_{\rm ext}/\gamma_{\rm r}$ is the angular velocity induced by an external torque, if present, and $\pmb{\Omega}_{\rm fl}(t)={\bf T}_{\rm fl}(t)/\gamma_{\rm r}$ a fluctuating angular velocity.  If the Janus particle carries a magnetic moment $\mu$, the external torque is given by ${\bf T}_{\rm ext}=\mu \, {\bf u}\times{\bf B}$ in the presence of an external magnetic field $\bf B$.

Since the particle is spherical, this equation does not depend on the particle position or the thermal state (if the external torque is spatially uniform).  Consequently, this stochastic equation is autonomous and it drives the direction $\bf u$ independently of what happens for translation and heat transfer.

\subsection{Radiative heat transfer}

The energy $E$ transferred during the time interval $[0,t]$ between the black-body radiation and the Janus particle is also ruled by a stochastic differential equation:
\be
\frac{dE}{dt} = \Phi_E + \Phi_{\rm sth} + \Phi_{\rm fl}(t) \, ,
\label{eq-E}
\ee
where $\Phi_E$ is the mean heat transfer~(\ref{Phi-E}), $\Phi_{\rm sth}$ a contribution from self-thermophoresis to be determined, and $\Phi_{\rm fl}(t)$ a fluctuating rate.  The fluctuating rate is a Gaussian white noise that can be expressed as
\be
\Phi_{\rm fl}(t) = \int_{C} \xi_{\phi}^{\rm s}({\bf r},t) \, d\Sigma
\ee
in terms of the noise $\xi_{\phi}^{\rm s}$ on the heating rate per unit area~(\ref{phi-s}) and the surface integral over the capped hemisphere $C$ that absorbs radiation.
Since this noise is characterized by Eqs.~(\ref{phi-noise}) and~(\ref{L-phi-phi}), its mean value is zero, $\langle\Phi_{\rm fl}(t)\rangle=0$, and its diffusivity is given by
\be
D_E = \lim_{t\to\infty} \frac{1}{2t} \Big\langle\left[\int_0^t \Phi_{\rm fl}(t')\, dt'\right]^2\Big\rangle = k_{\rm B} \int_C L_{\phi\phi}^{\rm s}\, d\Sigma \simeq 4\pi R^2 k_{\rm B}\, \epsilon\, \sigma\, (T_0^5+T_{\infty}^5)
\label{DE-gen}
\ee
in the approximation where the temperature of the absorbing layer uniformly takes the value $T_{\infty}$.
We note that the diffusivity $D_E$ is usually extremely small for thermal radiation, so that the radiative transfer of energy is practically non-fluctuating.

\subsection{Coupling in the linear regime close to equilibrium}

By the fundamental property of microreversibility, we can use the Onsager symmetry principle in order to determine $\Phi_{\rm sth}$ in the linear regime.  First, we identify the affinities or generalized thermodynamic forces as the mechanical and thermal affinities, respectively given by
\be
{\bf A}_{\rm mech} = \beta \, {\bf F}_{\rm ext} \qquad\mbox{and}\qquad A_E=(k_{\rm B}T_{\infty})^{-1}-(k_{\rm B}T_0)^{-1}\, .
\ee
These affinities are the parameters controlling how the system is driven away from equilibrium.

In the linear regime close to equilibrium, the thermal affinity and the radiation intensity~(\ref{intensity}) are proportional to the difference between the temperatures of the black-body radiation and the fluid at large distances from the particle, $\Delta T = T_0-T_{\infty}$.  Taking the temperature of the fluid as the reference temperature $T_{\infty}$, the thermal affinity, the mean heating rate~(\ref{Phi-E}), and the diffusivity~(\ref{DE-gen}) are approximately given by
\be
A_E \simeq \frac{\Delta T}{k_{\rm B} T_{\infty}^2} \, , \qquad \Phi_E \simeq 8\pi R^2 \epsilon\sigma T_{\infty}^3 \Delta T \, , \qquad\mbox{and}\qquad D_E\simeq 8\pi R^2 k_{\rm B}\, \epsilon\, \sigma\, T_{\infty}^5 \, .
\ee
In the linear regime, we thus have the approximation $\Phi_E = D_E  A_E$ as expected.

Gathering the variables, the affinities, and the noise terms in the four-dimensional vectors ${\bf X} = ({\bf R} ,  E )$, ${\bf A} = ({\bf A}_{\rm mech} ,  A_E  )$, and $\delta{\bf J}(t)=\left[{\bf V}_{\rm fl}(t),\Phi_{\rm fl}(t)\right]$, the coupled stochastic equations~(\ref{eq-R}) and~(\ref{eq-E}) can be expressed in the linear regime as
\be
\frac{d{\bf X}}{dt} = {\boldsymbol{\mathsf L}}\cdot{\bf A} + \delta{\bf J}(t)
\label{gen-eq}
\ee
 in terms of the matrix of linear response coefficients ${\boldsymbol{\mathsf L}}$ and the Gaussian white noise terms $\delta{\bf J}(t)$, satisfying $\langle\delta{\bf J}(t)\rangle=0$ and $\langle\delta{\bf J}(t)\, \delta{\bf J}(t')\rangle=({\boldsymbol{\mathsf L}}+{\boldsymbol{\mathsf L}}^{\rm T})\, \delta(t-t')$, where the superscript T denotes the transpose.  Since the variables $\bf R$ and $E$ are even under time reversal, the matrix ${\boldsymbol{\mathsf L}}$ must be symmetric in order to satisfy Onsager's reciprocal relations.  Consequently, we deduce that
\be
{\boldsymbol{\mathsf L}} =
\left(
\begin{array}{cc}
D_{\rm t} \, {\boldsymbol{\mathsf 1}} & \chi\, D_{E} \, {\bf u} \\
\chi\, D_{E} \, {\bf u} & D_{E}
\end{array}
\right),
\label{L}
\ee
where ${\boldsymbol{\mathsf 1}}$ is the $3\times 3$ identity matrix.
In order to satisfy the second law of thermodynamics, this matrix must be non-negative and, thus, the diffusivities satisfy $D_{\rm t}\ge 0$, $D_{E} \ge 0$, and $D_{\rm t}\ge \chi^2 D_{E}$.  As a consequence of Eqs.~(\ref{gen-eq}) and~(\ref{L}), the coupled stochastic differential equations for the position $\bf R$ and the energy $E$ are given by
\bea
\frac{d{\bf R}}{dt} &=& \chi \, \Phi_E \, {\bf u} + \beta D_{\rm t}\, {\bf F}_{\rm ext} + {\bf V}_{\rm fl}(t) \, , \label{eq-R-fin}\\
\frac{dE}{dt} &=& \Phi_E + \beta\chi D_{E}\, {\bf u}\cdot{\bf F}_{\rm ext} + \Phi_{\rm fl}(t) \, , \label{eq-E-fin}
\eea
with the fluctuating velocity ${\bf V}_{\rm fl}(t)$ and the fluctuating rate $\Phi_{\rm fl}(t)$ given by coupled Gaussian white noise processes characterized by
\bea
&&\langle {\bf V}_{\rm fl}(t)\rangle = 0 \ , \qquad \langle \Phi_{\rm fl}(t)\rangle = 0 \, , \label{noise1} \\
&&\langle {\bf V}_{\rm fl}(t)\, {\bf V}_{\rm fl}(t')\rangle = 2D_{\rm t} \, \delta(t-t') \, {\boldsymbol{\mathsf 1}} \, , \label{noise2}\\
&&\langle \Phi_{\rm fl}(t)\, \Phi_{\rm fl}(t')\rangle = 2D_{E} \, \delta(t-t') \, , \label{noise3}\\
&&\langle {\bf V}_{\rm fl}(t)\, \Phi_{\rm fl}(t')\rangle = 2\chi D_{E} \, {\bf u} \, \delta(t-t') \, . \label{noise4}
\eea
As required, Eq.~(\ref{eq-R-fin}) is identical to Eq.~(\ref{eq-R}) since $V_{\rm sth}=\chi \Phi_E$.  We emphasize that Eqs.~(\ref{eq-R-fin}) and~(\ref{eq-E-fin}) are coupled to Eq.~(\ref{eq-rot}) for rotation.

Although the self-thermophoretic velocity given by the first term in the right hand side of Eq.~(\ref{eq-R-fin}) is observed in the experiments of Refs.~\cite{JYS10,BC15,KCC16}, the reciprocal effect in principle described by the second term of Eq.~(\ref{eq-E-fin}) should be practically unobservable since the radiation diffusivity $D_E$ is usually very small.  This situation is in contrast to the mechanochemical coupling in self-diffusiophoresis, which leads to observable effects \cite{GK17,GK18a,HSGK18}.

\subsection{Bivariate fluctuation relation and entropy production}

As for self-diffusiophoresis in Ref.~\cite{GK17}, a fluctuation relation can here also be established for the stochastic process ruled by Eqs.~(\ref{eq-rot}), (\ref{eq-R-fin}), and (\ref{eq-E-fin}). The Fokker-Planck equation governing the time evolution of the probability density $p({\bf R},E,{\bf u};t)$ can be written as
\be
\partial_t p =-\partial_{\bf X}\cdot{\pmb{\cal J}} +D_{\rm r} \, \hat{\cal L}_{\rm r} p
\label{FP-eq}
\ee
with the current density
\be
\pmb{\cal J} = {\boldsymbol{\mathsf L}}\cdot{\bf A}\, p - {\boldsymbol{\mathsf L}}\cdot\partial_{\bf X}p
\label{FP-eq-J}
\ee
expressed in terms of the matrix~(\ref{L}) of linear response coefficients, the rotational diffusion coefficient $D_{\rm r}=(\beta\gamma_{\rm r})^{-1}$, and the operator
\be
\hat {\cal L}_{\rm r} p = \frac{1}{\sin\theta}\, \partial_{\theta} \left[ \sin\theta \, {\rm e}^{-\beta U_{\rm r}} \partial_{\theta}\left( {\rm e}^{\beta U_{\rm r}} p \right) \right] + \frac{1}{\sin^2\theta}\, \partial_{\varphi} \left[  {\rm e}^{-\beta U_{\rm r}} \partial_{\varphi}\left( {\rm e}^{\beta U_{\rm r}} p \right) \right] ,
\label{Lr}
\ee
where $U_{\rm r}= -\mu\, {\bf B}\cdot{\bf u}$ is a rotational energy associated with the external torque exerted on the particle by an external magnetic field $\bf B$.

Modifying the Fokker-Planck operator to include parameters counting the displacement and the energy transferred as carried out in Ref.~\cite{GK17}, we obtain the following fluctuation relation
\be
\frac{{\cal P}({\bf R},E,t)}{{\cal P}(-{\bf R},-E,t)} \simeq_{t\to\infty} \exp\left({\bf A}_{\rm mech}\cdot{\bf R} + A_E \, E \right)
\label{FT}
\ee
for the joint probability ${\cal P}({\bf R},E,t) = \int p({\bf R},E,{\bf u},t)\, d^2u $ of a displacement $\bf R$ of the Janus particle and the transfer of the radiation energy $E$ during the time interval $[0,t]$.

As a consequence, the thermodynamic entropy production
\be
\frac{1}{k_{\rm B}}\frac{d_{\rm i}S}{dt} ={\bf A}_{\rm mech}\cdot \langle{\bf \dot R}\rangle + A_E\langle\dot E\rangle \ge 0
\ee
is non-negative in accordance with the second law of thermodynamics.

\section{Numerical results}
\label{sec:num}

Here, the coupled Langevin equations~(\ref{eq-rot}), (\ref{eq-R-fin}), and~(\ref{eq-E-fin}) are simulated with the method described in Appendix~\ref{app:sim} in order to investigate the effects of self-thermophoresis on the stochastic motion of the Janus particle.  The results are presented in terms of the following dimensionless variables
\be
t_* \equiv D_{\rm r} t \, , \quad {\bf R}_* \equiv {\bf R} \sqrt{\frac{D_{\rm r}}{D_{\rm t}}} \, , \quad E_* \equiv E\sqrt{\frac{D_{\rm r}}{D_E}} \, , 
\ee
and parameters
\be
\Phi_{E*} \equiv \frac{\Phi_E}{\sqrt{D_{E}D_{\rm r}}}  \, , \quad \chi_* \equiv \chi \sqrt{\frac{D_E}{D_{\rm t}}} \, , \quad {\bf F}_{{\rm ext}*} \equiv \beta \, {\bf F}_{\rm ext} \sqrt{\frac{D_{\rm t}}{D_{\rm r}}} \, .
\ee

\subsection{Self-thermophoretic enhancement of diffusion}

In the absence of an external force and an external magnetic field, (${\bf F}_{\rm ext}=0$, ${\bf B}=0$), the self-thermophoretic propulsion combined with rotational diffusion without a preferred orientation has the effect of enhancing translational diffusion.  Indeed, integrating Eq.~(\ref{eq-R-fin}) with ${\bf F}_{\rm ext}=0$, we obtain the following effective translational diffusion coefficient
\be
D_{\rm t}^{\rm (eff)} \equiv \lim_{t\to\infty} \frac{1}{6t} \left\langle\left[ {\bf R}(t)-{\bf R}(0)\right]^2\right\rangle = D_{\rm t} + \frac{\chi^2 \Phi_E^2}{6 D_{\rm r}} ,
\label{Deff}
\ee
where the limit is reached beyond the time scale of rotational diffusion $t_{\rm rot \; diff}=(2D_{\rm r})^{-1}$.  Without self-thermophoretic activity ($\Phi_E=0$), the coefficient $D_{\rm t}$ for passive diffusion is recovered.  With self-thermophoresis, the effective diffusion coefficient increases quadratically with the heating rate $\Phi_E$ and the self-thermophoretic parameter~(\ref{chi}).

\begin{figure}[h]
\centerline{\scalebox{0.42}{\includegraphics{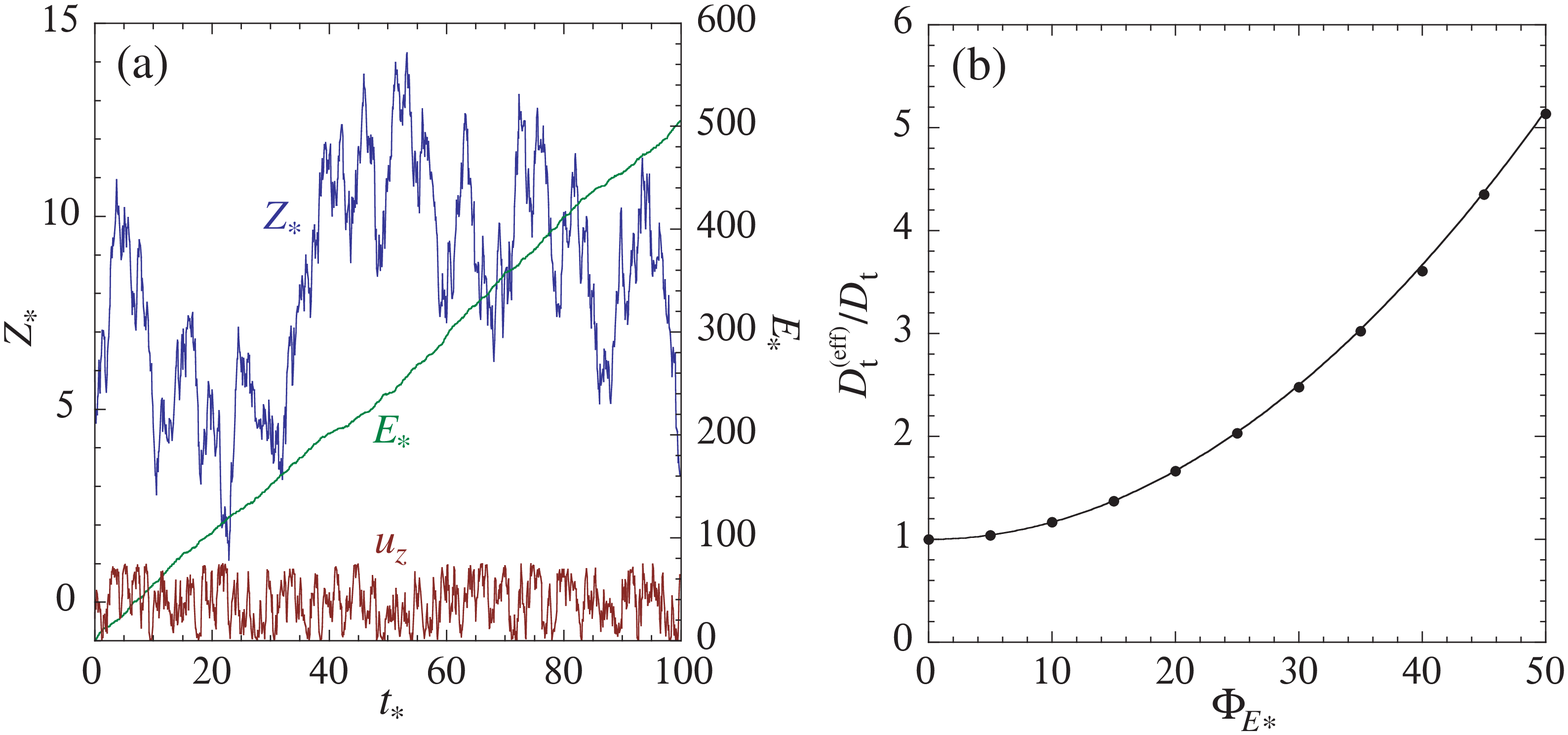}}}
\caption{Self-thermophoretic enhancement of diffusion with ${\bf F}_{\rm ext}=0$, ${\bf B}=0$, and $\chi_*=0.1$: (a) Example of stochastic time evolution for the position $Z_*$ in the $z$-direction, the transferred energy $E_*$ due to the radiative heating $\Phi_{E*}=5$, and the projection $u_z$ of the orientation.  The coupled Langevin equations are integrated with the time step $\Delta t_*=10^{-3}$. (b) The effective translational diffusion coefficient $D_{\rm t}^{\rm (eff)}$ (rescaled by $D_{\rm t}$) versus the dimensionless radiative heating rate $\Phi_{E*}$.  The dots show the results of a statistics over $10^5$ trajectories of $10^4$ time steps $\Delta t_*=10^{-2}$ obtained by the numerical integration of Eqs.~(\ref{eq-rot}), (\ref{eq-R-fin}), and~(\ref{eq-E-fin}). The line is the theoretical prediction of Eq.~(\ref{Deff}).}
\label{fig1}
\end{figure}

As shown in Fig.~\ref{fig1}, this behavior is confirmed by numerical simulations.  Figure~\ref{fig1}(a) depicts an example of stochastic time evolution for the random variables $Z_*$, $E_*$, and $u_z$, which are respectively the third component of position ${\bf R}_*$, the transferred energy, and the third component of the unit vector $\bf u$ giving the particle orientation.  Since there is no external magnetic field (${\bf B}=0$), the orientation $\bf u$ has uniform fluctuations in all directions, so that its projection $u_z$ has equal fluctuations in opposite directions.  Consequently, the displacement remains a random walk, although enhanced by the self-thermophoresis due to radiative heating.  The dependence of the effective translational diffusion coefficient on the radiative heating rate is shown in Fig.~\ref{fig1}(b), as computed by simulations (dots) and by Eq.~(\ref{Deff}) (line), which confirms the theoretical prediction of a enhancement of diffusion that is quadratic in the radiative intensity.

\subsection{Self-thermophoretic propulsion with external control of orientation}

In the presence of an external magnetic field ${\bf B}=(0,0,B)$ orienting the Janus particle in the $z$-direction but no external force ${\bf F}_{\rm ext}=0$, the self-thermophoretic propulsion has a persistent effect, causing the displacement of the Janus particle in the mean direction $\langle{\bf u}\rangle=(0,0,\langle u_z\rangle)$ with the mean velocity
\be
\langle \dot Z\rangle = \chi \, \Phi_E  \langle u_z\rangle\, , \qquad \mbox{where} \qquad \langle u_z\rangle = \coth(\beta\mu B) - \frac{1}{\beta\mu B} \, .
\label{Vz}
\ee

\begin{figure}[h]
\centerline{\scalebox{0.42}{\includegraphics{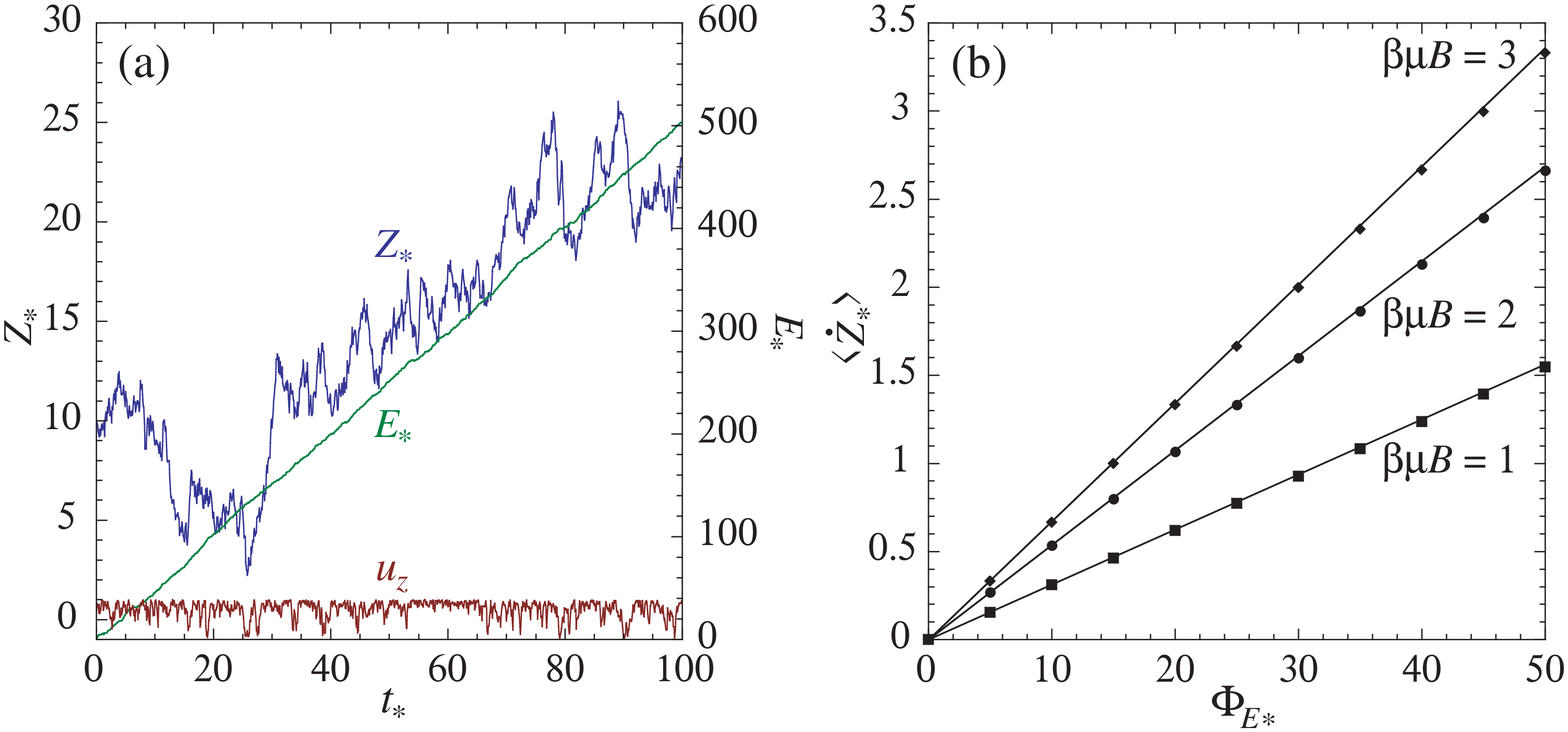}}}
\caption{Self-thermophoretic propulsion by external orientation with ${\bf F}_{\rm ext}=0$, ${\bf B}=(0,0,B)$, and $\chi_*=0.1$: (a) Example of stochastic time evolution for the position $Z_*$ in the $z$-direction, the transferred energy $E_*$ due to the radiative heating $\Phi_{E*}=5$, and the projection $u_z$ of the orientation, in the external magnetic field $\beta\mu B=2$.  The coupled Langevin equations are integrated with the time step $\Delta t_*=10^{-3}$. (b) The dimensionless mean velocity component $\langle\dot{Z}_*\rangle$ in the $z$-direction of the external magnetic field versus the dimensionless radiative heating rate $\Phi_{E*}$ for the values $\beta\mu B=1, 2, 3$.  The dots show the results of a statistics over $10^5$ trajectories of $10^4$ time steps $\Delta t_*=10^{-2}$ obtained by the numerical integration of Eqs.~(\ref{eq-rot}), (\ref{eq-R-fin}), and~(\ref{eq-E-fin}). The lines give the theoretical predictions of Eq.~(\ref{Vz}).}
\label{fig2}
\end{figure}

Figure~\ref{fig2}(a) shows an example of stochastic time evolution of the random variables $Z_*$, $E_*$, and $u_z$.  Because of the presence of the external magnetic field ${\bf B}=(0,0,B)$, the unit vector $\bf u$ is preferentially oriented in the $z$-direction, which is seen at the bottom of Fig.~\ref{fig2}(a) where $u_z$ is observed to fluctuate with a higher probability near $u_z=+1$ than near $u_z=-1$.  As a consequence, the Janus particle undergoes a random drift in the $z$-direction due to self-thermophoresis by radiative heating. Figure~\ref{fig2}(b) confirms the prediction of Eq.~(\ref{Vz}) that the mean velocity of the Janus particle increases linearly with the radiative heating rate $\Phi_{E*}$ and with the magnitude of the rescaled external magnetic field $\beta\mu B$, which enhances the orientation of the Janus particle and thus its propulsion.

\subsection{Motor stall by an external force with controlled orientation}

In the presence of both an external force ${\bf F}_{\rm ext}=(0,0,F)$ and an external magnetic field ${\bf B}=(0,0,B)$, the Janus particle is not only oriented in the $z$-direction, but is also driven by the external force.  The unit vector giving the orientation of the Janus particle is polarized on average in the direction of the external magnetic field, $\langle{\bf u}\rangle=(0,0,\langle u_z\rangle)$, where $\langle u_z\rangle$ is given in Eq.~(\ref{Vz}).  Therefore, the mean values of Eqs.~(\ref{eq-R-fin}) and~~(\ref{eq-E-fin}) give
\bea
&&\langle \dot X\rangle = \langle \dot Y\rangle = 0 \, , \label{Av-eq1-Fext}\\
&&\langle \dot Z\rangle = \chi \, \Phi_E  \langle u_z\rangle + \beta  D_{\rm t} F \, , \label{Av-eq2-Fext}\\
&&\langle \dot E\rangle = \Phi_E + \beta  \chi  D_E  \langle u_z\rangle  F \, . \label{Av-eq3-Fext}
\eea
As expected, the mean velocity in the direction of the magnetic field now has two contributions: those from self-thermophoresis and the external force.  Consequently, the motor stalls at the critical value $F_{\rm stall}=-\chi \, \Phi_E  \langle u_z\rangle/(\beta  D_{\rm t})$ of the external force.  Furthermore, the mean value of the radiative heating rate also depends on the external force because of the Onsager reciprocal relations.

\begin{figure}[h]
\centerline{\scalebox{0.52}{\includegraphics{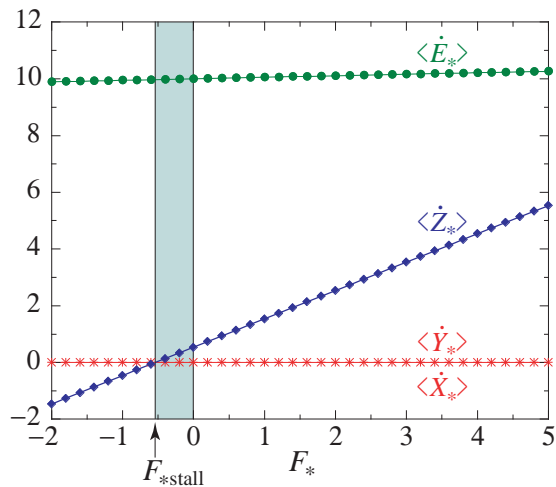}}}
\caption{Self-thermophoretic propulsion with an applied external force and magnetic field to control the orientation with ${\bf F}_{\rm ext}=(0,0,F)$, ${\bf B}=(0,0,B)$, $\beta\mu B=2$, $\Phi_{E*}=5$, and $\chi_*=0.1$:  The dimensionless mean velocity components $(\langle\dot{X}_*\rangle,\langle\dot{Y}_*\rangle,\langle\dot{Z}_*\rangle)$ and the mean radiative heating rate $\langle\dot{E}_*\rangle$ versus the dimensionless external force $F_*=\beta F \sqrt{D_{\rm t}/D_{\rm r}}$ are plotted in the figure.  The dots show the results of a statistics over $10^5$ trajectories of $10^4$ time steps $\Delta t_*=10^{-2}$ obtained by the numerical integration of Eqs.~(\ref{eq-rot}), (\ref{eq-R-fin}), and~(\ref{eq-E-fin}). The lines give the theoretical predictions of Eqs.~(\ref{Av-eq1-Fext})-(\ref{Av-eq3-Fext}).  The force interval where radiative heating performs mechanical work is shown in green.}
\label{fig3}
\end{figure}

Figure~\ref{fig3} confirms these theoretical expectations by numerical simulations.  In this figure, we observe that the mean velocity in the direction of the magnetic field indeed increases with the external force and motor stall occurs at the dimensionless value $F_{*\, {\rm stall}}=-0.537$, as predicted by Eq.~(\ref{Av-eq2-Fext}).  In the interval $F_{*\,{\rm stall}}<F_*<0$, radiative heating performs mechanical work on the Janus particle due to the self-thermophoretic coupling, since the mean velocity $\langle\dot{Z}_*\rangle$ is positive although the external force $F_*$ is negative.  In addition, the radiative heating rate $\langle\dot{E}_*\rangle$ also increases with the external force due to the reciprocal effect of the external force back onto radiative heating; however, as discussed earlier, this reciprocal effect is very small for  self-thermophoresis.

\section{Summary and perspectives}\label{sec:conc}

A thermodynamically consistent derivation of Langevin equations for the self-thermophoretic dynamics of Janus motors partially coated with an absorbing layer that is heated by a radiation field was given in this paper, based on fluctuating bulk phase and surface equations. In particular, the surface equations for the surface pressure tensor and heat flux vector are expressed in terms of the slip velocity and surface temperature gradient and are constructed to be consistent with microscopic reversibility. The resulting full Langevin equations allow one to obtain general expressions for the thermophoretic force and torque on the Janus colloid for partial slip boundary conditions. The overdamped Langevin equations for the colloid displacement and radiative heat transfer, which satisfy Onsager symmetry relations as a consequence of microscopic reversibility, show the presence of a reciprocal effect where an external force changes the radiative heat transfer. It was also shown that although this contribution is necessarily present for thermodynamic consistency its magnitude is small, in contrast to self-diffusiophoresis. An explicit expression for the self-thermophoretic velocity for general slip boundary conditions is provided and compared with the analogous expression for thermophoretic motion in an external temperature gradient. The nonequilibrium fluctuation formula provides a relation between the probability density of the Janus particle displacement and the radiation energy transfer during the time interval $[0,t]$ and the mechanical and thermal affinities that characterize the nonequilibrium system state.

The results obtained in this paper should provide the basis for the analysis of experimental and simulation studies of self-thermophoretic motion.

\section*{Acknowledgments}

Financial support from the International Solvay Institutes for Physics and Chemistry, the Universit\'e libre de Bruxelles (ULB), the Fonds de la Recherche Scientifique~-~FNRS under the Grant PDR~T.0094.16 for the project ``SYMSTATPHYS", and the Natural Sciences and Engineering Research Council of Canada is acknowledged.

\appendix
\section{Thermophoretic velocity of a spherical particle}
\label{app:therm-vel}

In this appendix, we obtain the thermophoretic velocity of a spherical particle moving in a temperature gradient without radiative heating.  In this case, thermophoresis manifests itself without the need of heterogeneity at the surface of the particle as for self-thermophoresis; thus, the particle is taken to be spherical with uniform properties at its surface.

In the thermophoretic case, the temperature profile is obtained by solving the Laplace equation~(\ref{Laplace-eq}) with the following boundary conditions:
\bea
&&T^+(R,\theta) = T^-(R,\theta) \, , \label{T-bc1-th}\\
&&\kappa^+\partial_rT^+(R,\theta) =\kappa^- \partial_rT^-(R,\theta) \, , \label{T-bc2-th}\\
&&\lim_{r\to\infty} \pmb{\nabla}T^+(r,\theta) = g\, {\bf 1}_z\, , \label{T-bc3-th}
\eea
instead of the boundary conditions~(\ref{T-bc1})-(\ref{T-bc3}).  Here, we consider spherical coordinates $(r,\theta,\varphi)$ with $\theta=0$ along the axis of the temperature gradient ${\bf g}=g\, {\bf 1}_z$.
Solving this problem, we get the following temperature field:
\bea
&& r>R: \qquad T^+(r,\theta) = T_0 + \left[ 1 + \frac{\kappa^+-\kappa^-}{2\kappa^++\kappa^-}\left(\frac{R}{r}\right)^3\right] {\bf g}\cdot{\bf r}  \, , \\
&& r<R: \qquad T^-(r,\theta) = T_0 +  \frac{3\kappa^+}{2\kappa^++\kappa^-}\, {\bf g}\cdot{\bf r}\, .
\eea
Here, $T_0$ denotes the temperature at the center of the particle.
At the surface of the spherical particle, the temperature has the profile
\be
T^+(R,\theta) = T^-(R,\theta) = T_0 +  \frac{3\kappa^+}{2\kappa^++\kappa^-}\, Rg\cos\theta \, .
\label{T-surf-th}
\ee
Consequently, the thermophoretic velocity is given by
\be
{\bf V}_{\rm th} =  \frac{b_q}{1+2b/R} \, \frac{2\kappa^+ }{2\kappa^++\kappa^-}\, {\bf g} \, ,
\label{Vth}
\ee
where ${\bf g}=\pmb{\nabla}T^+$ is the temperature gradient at large distances from the particle, which confirms the known expression for this quantity~\cite{BZW14}.

As expected, both thermophoretic velocity (\ref{Vth}) and the self-thermophoretic velocity (\ref{V-th}) are proportional to the thermophoretic constant $b_q$. However, here, the particle is driven away from equilibrium by the temperature gradient instead of the absorbed radiation intensity.

If a colloidal suspension is dilute and subjected to a temperature gradient, the current density of colloids can be expressed as
\be
{\bf J} = -D \pmb{\nabla} n - n \, D_T \, \pmb{\nabla}T = -D\left( \pmb{\nabla} n + n \, S_T \, \pmb{\nabla}T\right)
\label{Crnt}
\ee
in terms of the colloidal number density $n$, the mutual diffusion coefficient $D$, the thermodiffusion coefficient $D_T$, or equivalently the Soret coefficient $S_T\equiv D_T/D$ \cite{GM84,W07,PP08,LYR12,BBPE18}.  The diffusion coefficient is given by $D=D_{\rm t}=(\beta\gamma_{\rm t})^{-1}$ with the translational friction coefficient~(\ref{t-friction}).  The current density~(\ref{Crnt}) can also be expressed as ${\bf J} = -D \pmb{\nabla} n + n {\bf V}_{\rm th}$ in terms of the thermophoretic velocity~(\ref{Vth}).  Accordingly, the thermodiffusion coefficient is here given by
\be
D_T = - \frac{b_q}{1+2b/R} \, \frac{2\kappa^+ }{2\kappa^++\kappa^-}
\ee
and the Soret coefficient by
\be
S_T = \frac{D_T}{D}= - \frac{6\pi\eta R}{1+3b/R} \, \frac{b_q}{k_{\rm B}T} \, \frac{2\kappa^+ }{2\kappa^++\kappa^-} \, .
\ee

\section{Method for simulating the stochastic motion}
\label{app:sim}

The rotational motion described by Eq.~(\ref{eq-rot}) can be simulated by the method of quaternions \citep{IBdO15,DBD15} and the fluctuating torque $\pmb{\Omega}_{\rm fl}(t)=\sqrt{2D_{\rm r}}\, \pmb{\xi}_{\rm r}(t)$ is determined by three independent Gaussian noise terms such that $\langle\pmb{\xi}_{\rm r}(t)\rangle = 0$ and $\langle\pmb{\xi}_{\rm r}(t)\,\pmb{\xi}_{\rm r}(t')\rangle = \delta(t-t')\, {\boldsymbol{\mathsf 1}}$ with the rotational diffusion coefficient $D_{\rm r}=(\beta\gamma_{\rm r})^{-1}$.  Also, the coupled noise terms~(\ref{noise1})-(\ref{noise4}) of the overdamped Langevin equations~(\ref{eq-R-fin}) and~(\ref{eq-E-fin}) can be simulated according to
\bea
&&{\bf V}_{\rm fl}(t) = \sqrt{2D_{\rm t}} \xi_1(t)  {\bf u}_1 + \sqrt{2D_{\rm t}} \xi_2(t)  {\bf u}_2 +\left[\sqrt{D_{\rm t}\lambda_+} \xi_3(t) - \varsigma\sqrt{D_{\rm t}\lambda_-} \xi_4(t)\right]  {\bf u}_3 , \qquad \label{V_fl_decomp}\\
&&\Phi_{\rm fl}(t) = \varsigma\sqrt{D_{E}\lambda_+} \xi_3(t) + \sqrt{D_{E}\lambda_-} \xi_4(t) ,
\label{W_fl_decomp}
\eea
in terms of other independent Gaussian white noise terms satisfying $\langle\xi_i(t)\rangle = 0$ and $\langle\xi_i(t)\, \xi_j(t')\rangle = \delta(t-t') \, \delta_{ij}$, the parameters $\lambda_{\pm}=1\pm\sqrt{\chi^2D_{E}/D_{\rm t}}$ and $\varsigma=\chi/\vert\chi\vert$, and the unit vectors $\{{\bf u}_1,{\bf u}_2,{\bf u}_3={\bf u}\}$ attached to the frame of the Janus particle.  The unit vector ${\bf u}_3={\bf u}$ is oriented along the particle axis pointing from the transparent towards the absorbing hemisphere, while ${\bf u}_1$ and ${\bf u}_2$ are perpendicular to the axis.

Using Eqs.~(\ref{V_fl_decomp}) and~(\ref{W_fl_decomp}), we have that
\bea
{\bf u}(t)\cdot{\bf V}_{\rm fl}(t) &=& \sqrt{D_{\rm t}\lambda_+}\, \xi_3(t) -\ \ \varsigma\sqrt{D_{\rm t}\lambda_-}\, \xi_4(t)  \, , \\
\Phi_{\rm fl}(t) &=& \varsigma\sqrt{D_{E}\lambda_+}\, \xi_3(t) + \sqrt{D_{E}\lambda_-}\, \xi_4(t) \, .
\eea
As a consequence, we get
\bea
&&\langle{\bf u}(t)\cdot {\bf V}_{\rm fl}(t)\rangle = 0 \ , \qquad \langle \Phi_{\rm fl}(t)\rangle = 0 \, , \label{noise1b} \\
&&\langle {\bf u}(t)\cdot {\bf V}_{\rm fl}(t)\, {\bf u}(t')\cdot {\bf V}_{\rm fl}(t')\rangle = 2D_{\rm t} \, \delta(t-t') \, \label{noise2b}\\
&&\langle \Phi_{\rm fl}(t)\, \Phi_{\rm fl}(t')\rangle = 2D_{E} \, \delta(t-t') \, , \label{noise3b}\\
&&\langle {\bf u}(t)\cdot{\bf V}_{\rm fl}(t)\, \Phi_{\rm fl}(t')\rangle = 2\chi D_{E} \, \delta(t-t') \, . \label{noise4b}
\eea
At equilibrium where $\Phi_E=0$ and ${\bf F}_{\rm ext}=0$, we have that ${\bf \dot R}(t)={\bf V}_{\rm fl}(t)$ and $\dot E(t)= \Phi_{\rm fl}(t)$, which can be substituted into Eqs.~(\ref{noise2b})-(\ref{noise4b}).  Therefore, integrating Eqs.~(\ref{noise2b})-(\ref{noise4b}) over time under equilibrium conditions gives the following Green-Kubo formulae,
\bea
&& D_{\rm t} = \frac{1}{2} \int_{-\infty}^{+\infty} \langle {\bf u}(t)\cdot{\bf \dot R}(t) \,  {\bf u}(0)\cdot{\bf \dot R}(0) \rangle_{\rm eq} \, dt \, , \label{GK1}\\
&& D_{E} = \frac{1}{2} \int_{-\infty}^{+\infty} \langle {\dot E}(t) \,  {\dot E}(0) \rangle_{\rm eq} \, dt \, , \label{GK2}\\
&& \chi\, D_{E} = \frac{1}{2} \int_{-\infty}^{+\infty} \langle {\bf u}(t)\cdot{\bf \dot R}(t) \,  {\dot E}(0) \rangle_{\rm eq} \, dt = \frac{1}{2} \int_{-\infty}^{+\infty} \langle {\dot E}(t) \,  {\bf u}(0)\cdot{\bf \dot R}(0) \rangle_{\rm eq} \, dt \, , \label{GK3}
\eea
for the diffusivities and the self-thermophoretic parameter $\chi$.


\end{document}